%% file: SPeris-15thQuark_Confinement.tex
\documentclass{webofc}
\usepackage[varg]{txfonts}   
\input defs.tex

\begin{document}
\title{\vspace{0cm}$\alpha_s$ from an improved $\tau$ vector isovector spectral function}
%
%

\author{\firstname{Diogo} \lastname{Boito}\inst{1}
\and
        \firstname{Maarten} \lastname{Golterman}\inst{2}
        \and
        \firstname{Kim} \lastname{Maltman}\inst{3}
        \and
        \firstname{Santiago} \lastname{Peris}\inst{4}\fnsep\thanks{peris@ifae.es}\fnsep\thanks{Speaker}
             \and
        \firstname{Marcus V.} \lastname{Rodrigues}\inst{5}
             \and
        \firstname{Wilder} \lastname{Schaaf}\inst{6}
}

\institute{Instituto de F\'{\i}sica de S{\~a}o Carlos, Universidade de S{\~a}o Paulo, \\
CP 369, 13570-970, S{\~a}o Carlos, SP, Brazil
\and
          Department of Physics and Astronomy, San Francisco State University,\\
San Francisco, CA 94132, USA
\and
          Department of Mathematics and Statistics,
York University\\  Toronto, ON Canada M3J~1P3
\and
Department of Physics and IFAE-BIST, Universitat Aut\`onoma de Barcelona\\
E-08193 Bellaterra, Barcelona, Spain
\and
Deutsches Elektronen-Synchrotron (DESY),
Notkestra{\ss}e 85, 22607 Hamburg, Germany
\and
Department of Physics, University of Washington, Seattle, WA 98195-1560, USA
}

\abstract{
After discussing difficulties in determining $\alpha_s$ from tau decay due
to the existence of Duality Violations and the associated asymptotic nature
of the OPE, we describe a new determination based on an improved vector
isovector spectral function, now based solely on experimental input, obtained
by (i) combining ALEPH and OPAL results for $2\pi +4\pi$ and (ii) replacing
$K^-K^0$ and higher-multiplicity exclusive-mode contributions, both
previously estimated using Monte Carlo, with new experimental BaBar results
for $K^-K^0$ and results implied by $e^+ e^-$ cross sections and CVC for
the higher-multiplicity modes. We find $\alpha_s(m_\tau)=0.3077\pm 0.0075$,
which corresponds to $\alpha_s(m_Z)=0.1171\pm 0.0010$. Finally, we
comment on some of the shortcomings in the criticism of our approach
by Pich and Rodriguez-Sanchez.
}

\maketitle

It has been clear since the pioneering work of
Ref.~\cite{Shankar:1977ap} (see also Ref.~\cite{Braaten:1991qm}
for other pre-1992 references and a then-up-to-date implementation
of the approach) that Finite Energy Sum Rules (FESRs) provide a
potentially useful tool for extracting $\alpha_s$ from hadronic $\tau$
decay data. Subsequent increases in the precision of both the experimental
data \cite{OPAL:1998rrm,Davier:2013sfa} and the order to which the
dominant perturbative contribution is known \cite{Baikov:2008jh} have
significantly reduced the resulting uncertainty on $\alpha_s$, to
the extent that assumptions/approximations which might have
been reasonable in 1992 now need to be revisited.

In our context, the FESR consists of the identity
\beq
\lbl{FESR}
\underbrace{\int_{s_{th}}^{s_0}\frac{ds}{s_0}\ w\left(\frac{s}{s_0}\right)
\frac{1}{\p}\mathrm{Im}\P(s)}_{I_w^{exp}(s_0)}\  =\
\underbrace{-\frac{1}{2i\p}\oint_{|z|=s_0}\frac{dz}{s_0}\ \hat{w}
\left(\frac{z}{s_0}\right)\ D(z)}_{I_w^{th}(s_0)}\ ,
\eeq
with $\P(s)$ the correlator associated with the $I=1$ vector
current $\bar{u}\g^\m d$, $w$ and $\hat{w}$ polynomials related
by an integration by parts, and $D(z)$ the Adler function,
$D(z)=-z\frac{d\Pi(z)}{dz}$.\footnote{In general, the axial
current can also be considered.}

At large enough $s_0$, it is reasonable to approximate the RHS using
the Operator Product Expansion (OPE)\footnote{In what follows, we
consider the perturbative series as the contribution from the
identity operator.} and extract
$\alpha_s$ using experimental data for $\mathrm{Im}\, \Pi (z)$ as input on the LHS. The presence of a cut in $\Pi (z)$,
however, ensures that the OPE is at best an asymptotic expansion in
$1/z$.\footnote{A convergent expansion in inverse powers of $z$ must
have a disc of convergence.} The RHS must, then, in general, contain
an additional, non-OPE ("duality violating", or DV) contribution to
compensate for the lack of convergence on the circle $\vert z\vert =s_0$
\cite{Cata:2008ru}. Explicitly,
     \beq
\lbl{FESRDV}
-\frac{1}{2i\p}\oint_{|z|=s_0}\frac{dz}{s_0}
\hat{w}\left(\frac{z}{s_0}\right)\ D(z)=-\frac{1}{2i\p}
\oint_{|z|=s_0}\frac{dz}{s_0}\ \hat{w}\left(\frac{z}{s_0}\right)
\ D_{OPE}(z)-\int_{s_0}^\infty \frac{ds}{s_0}w\left(\frac{s}{s_0}\right)
\mathrm{Im}\P_{DV}(s) \ ,
\eeq
where the DV contribution, $\mathrm{Im}\, \Pi_{DV}(s)$, accounts for the
oscillatory behavior seen at lower $s$ in the spectrum, before
perturbative dominance sets in. This oscillatory behavior, which
cannot be accounted for by any power behavior in the OPE, led
Ref.~\cite{Poggio:1975af} to conclude that the OPE should not be
used "too close" to the cut. For a number of years this
stricture was implemented by "pinching", i.e., by choosing for
$\hat{w}(s/s_0)$ polynomials having a higher order zero at $s=s_0$,
thus suppressing contributions from near the cut. One, however, faces
two potential problems. First, it is not known {\it a priori} how much
pinching is needed for a determination of $\alpha_s$ free from DV
contamination. Second, a polynomial with a higher-degree zero
is necessarily higher degree in $s$, and generates higher-dimension, $D$,
OPE contributions on the RHS of Eq.~(\ref{FESRDV}). This is potentially
problematic, not only because the relevant higher-$D$ condensates are not
known, but also because an asymptotic expansion like the OPE ceases to
be valid at high orders. This leads us to our first message:

\vspace{.1cm}
\emph{It is not possible to simultaneously suppress the contribution from
DVs and high-order condensates. One should restrict oneself to low orders
of the OPE, but in a consistent manner.}
\vspace{.1cm}

Using a high-degree polynomial with strong pinching, but truncating the OPE
at low $D$, when unsuppressed higher-$D$ contributions are, in principle,
present, is thus a dangerous practice and leads, not surprisingly to
inconsistencies \cite{Boito:2019iwh}. We comment further
on this practice, which we refer to as the "truncated OPE"  (tOPE) approach
\cite{Pich:2016bdg,Davier:2013sfa,Ayala:2022cxo}, in Sec. \ref{correlations}.

The above discussion makes it clear it is not safe to ignore
DV contributions without further investigation. While no first-principles
derivation of the form of $\mathrm{Im}\, \Pi_{DV}(s)$ exists, some of its general
properties are known. As for the asymptotic expansion in powers of
the coupling $g^2$, where terms missed in the expansion are known to behave
as $e^{-const/g^2}$,\footnote{Recall the case of renormalons and the
perturbative series.} so terms missed in the OPE expansion of $\mathrm{Im}\,\Pi (s)$
are expected to behave as $e^{-const\cdot s}\times (\mathrm{oscillation})$. This
expectation was confirmed in Refs.~\cite{Boito:2017cnp,Peris:2021jap},
where the combination of a Regge-like spectrum ($M_n^2\sim n$) asymptotically and a
stringy relation ($\Gamma_n\sim M_n/N_c$) between resonance masses
and widths at large (but finite) number of colors, $N_c$, and large
resonance excitation number, $n$,\footnote{These are properties of QCD in
2 dimensions in the large-$N_c$ limit and also born out phenomenologically
in the real world \cite{Shifman:2007xn,Blok:1997hs,Masjuan:2012gc}} was
shown to lead to the large-$s$ expectation
\beq
\lbl{DVs}
\frac{1}{\p}\mathrm{Im}\P_{DV}(s)= e^{-\d-\g s} \sin\Bigg(\a + \b s+ \mathcal{O}\left(\log s\right) \Bigg)
\left( 1+ \mathcal{O}\left( \frac{1}{N_c},\frac{1}{\log s},
\frac{1}{s}\right) \right)\ .
\eeq
We will use Eq.~(\ref{DVs}) in conjunction with Eq.~(\ref{FESRDV}),
and comment on the impact of possible subleading corrections in
Sec.~\ref{duality}.

\section{Analysis}
\label{analysis}

A determination of $\alpha_s(m_\tau )$ requires one to specify (i) the
treatment of the perturbative series; (ii) the choice of weight $w(s)$ in
(\ref{FESR}) and whether the tOPE is used; (iii) whether
or not DVs are included; and (iv) last, but not least, the experimental data
to be used.

The perturbative series is currently known to order
$\alpha_s^4$~\cite{Baikov:2008jh}, with a generous guesstimate of the
$\alpha_s^5$ coefficient. The existence of the contour in Eq.~(\ref{FESR}),
however, leaves open the possibility of playing with the dependence on the
renomalization scale $\m$. Two major choices have been most popular. In
Contour Improved Perturbation Theory (CIPT)
\cite{Pivovarov:1991rh,LeDiberder:1992jjr}, the perturbative scale $\m^2$
is set equal to the variable, complex, local value $s_0 e^{i \phi}$ at
each point on the contour. In Fixed Order Perturbation Theory (FOPT), in
contrast, the same fixed-scale choice $\m^2=s_0$ is used at all points on
the contour. With $s_0>0$ a euclidean quantity, the FOPT choice naturally
matches the usual $\overline{\mathrm{MS}}$ scheme. For many years
$\alpha_s(m_\tau )$ has been obtained by averaging CIPT and FOPT results,
with a systematic error component given by half their difference. Recently,
however, Hoang and Regner \cite{Hoang:2020mkw} have shown that CIPT gives rise
to a perturbative series incompatible with the OPE. Pending a complete analysis where this fundamental flaw is fixed \cite{Benitez-Rathgeb:2022yqb,Benitez-Rathgeb:2022hfj},
all determinations of $\alpha_s(m_\t)$ employing CIPT to date should be
discarded. In what follows, therefore, we use only FOPT. This leads
us to our second message:

\vspace{.1cm}
\emph{Previous CIPT-based analyses should be discarded. Unless CIPT is
properly mended, averaging CIPT and FOPT values for $\alpha_s(m_\tau )$
should be avoided.}
\vspace{0.1cm}

We next turn to the choice of weights for use in Eq.~(\ref{FESR}). In
Ref.~\cite{Boito:2020xli} we chose a set of 4 linearly independent
polynomials of degree $\leq 4$ having no linear term: $w_0(y)=1,\, w_2(y)=
1-y^2,\, w_3(y)= (1-y)^2 (1+2y),\, w_4(y)= (1-y^2)^2$. The linear term is
avoided because renormalon model analyses show FESRs with such a term to be
problematic \cite{Beneke:2012vb}. This choice has the advantage of variable
sensitivity to DVs: $w_0$ has no pinching, while $w_{2,3,4}$ are
singly, doubly and doubly pinched, respectively. It also produces variable
sensitivity to the OPE condensates: up to $\alpha_s$-suppressed logarithmic
corrections, the $w_0$ FESR is free of condensate contributions and sensitive
only to perturbation theory while the $w_{2,3,4}$ FESRs receive
contributions also from the $D=6$ condensate, both $D=6$ and $8$ condensates,
and both $D=6$ and $10$ condensates, respectively, with $D=6$ contributions
in the ratios $1:3:2$. The consistency (or lack thereof) of results for this
condensate obtained from combined $w_0$ and $w_2$, combined $w_0$ and $w_3$
and combined $w_0$ and $w_4$ analyses provides a non-trivial check on our
analysis framework. Fits to Eq.~(\ref{FESRDV}) determine, in addition to
$\alpha_s$, the relevant OPE condensates and the DV parameters $\a, \b,\g$
and $\d$, also the value of $s$,
$s_{min}$, above which the parametrization
(\ref{DVs}) admits fits compatible with the underlying weighted spectral
integrals.

\begin{figure}[h]
\centering
\includegraphics[width=6cm,clip]{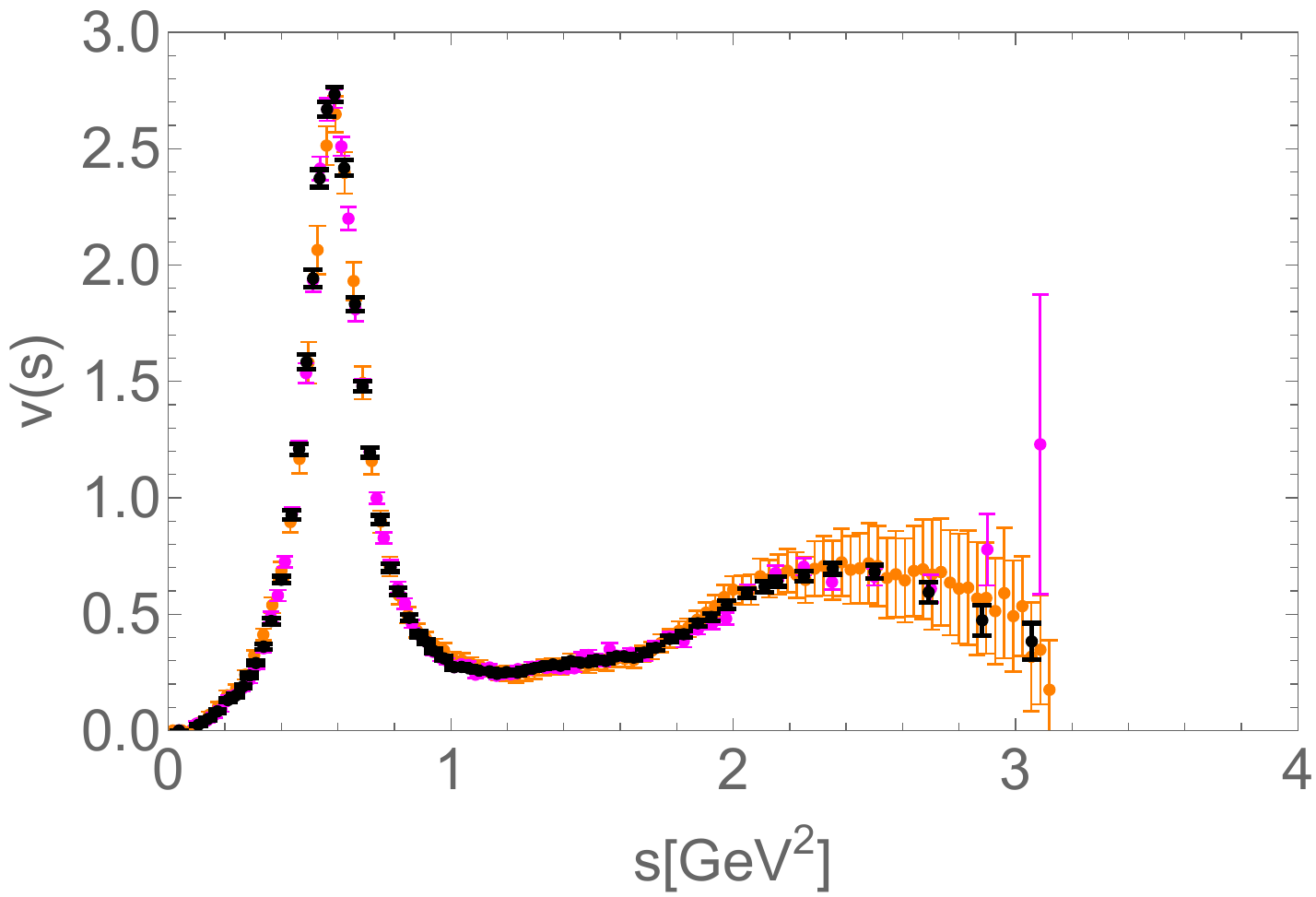}
\caption{ALEPH (magenta) and OPAL (orange) vector spectral functions,
$\mathrm{v}(s)= 2\p \mathrm{Im}\P(s)$, together with the improved
combined result of Ref.~\cite{Boito:2020xli} described in the text (black). }
\label{Fig. 1}       
\end{figure}

We now turn to the experimental data. Two major compilations of $\t$ data
have been used in previous $\a_s(m_\t)$ \cite{OPAL:1998rrm,Davier:2013sfa}
determinations. While this data is on average very precise, contributions
from $K\bar{K}$ and several of the multiparticle states relevant in the
higher-$s$ region were based on Monte Carlo rather than experimental
data. Fortunately, (i) BaBar results now exist for the $\t\to K^- K_s\n_\t$
distribution \cite{BaBar:2018qry}, and (ii) $e^+e^-$ data\footnote{See
Ref.~\cite{Boito:2020xli} for a full list of references.} combined with the
Conserved Vector Current relation allow the multiparticle-mode contributions
to be replaced with experimental input.\footnote{Isospin-breaking corrections
to the CVC relation are safely negligible for contributions from this part of
the spectrum.} Adding the contributions obtained using updated branching
fractions and combining ALEPH and OPAL $2\pi +4\pi$ results following the
algorithm of Ref.~\cite{Keshavarzi:2018mgv}, used for the R-data analysis
of the muon $g-2$, we obtained the updated $I=1$ vector spectral function
shown in Fig.~\ref{Fig. 1} \cite{Boito:2020xli}, which also shows the
corresponding original ALEPH and OPAL results. The gain in precision near
the end point of the spectrum is striking.

\section{Results}
\label{results}

\begin{figure}[h]
\centering

{\includegraphics[width=6cm]{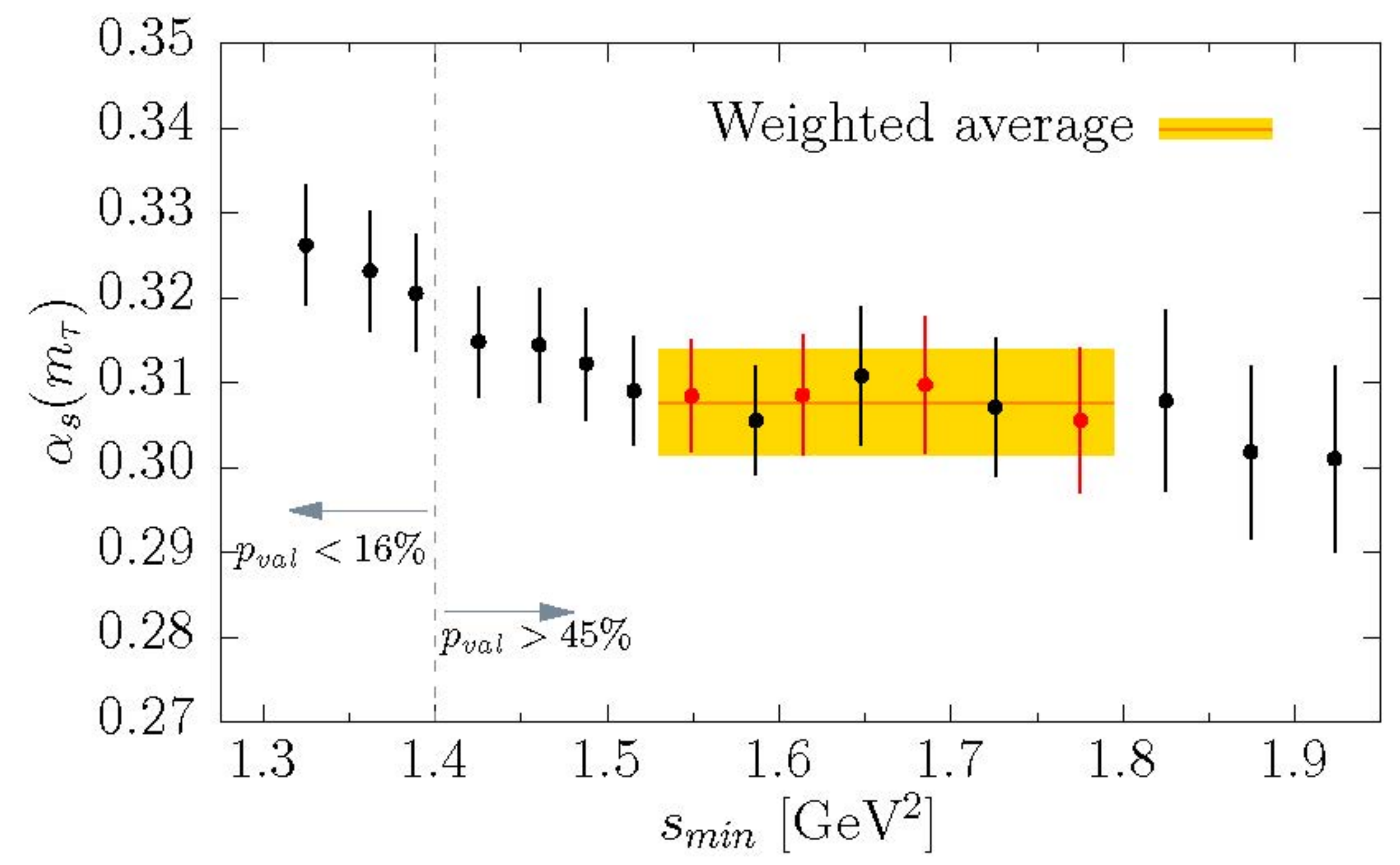}}
{\includegraphics[width=6.5cm]{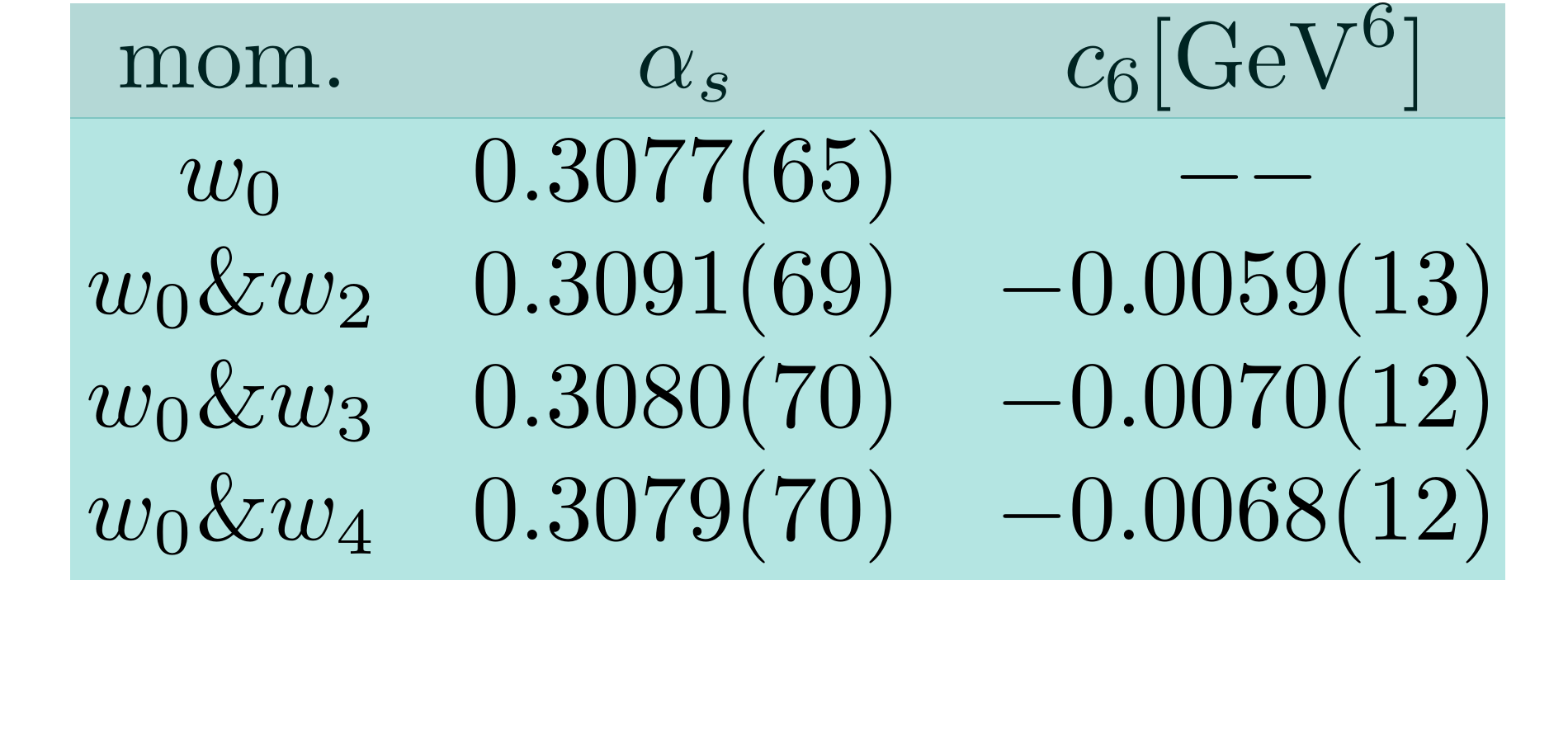}}

\caption{Left panel: Single-weight $w_0=1$ fit results for $\alpha_s(m_\tau )$
vs. $s_{min}$. P-values are $<16\%$ to the left of the dashed vertical line
and $>45 \%$ to the right. Right panel: results for $\alpha_s(m_\tau )$ and the
$D=6$ condensate, $c_6$, from fits using the weights shown in the first
column.}
\label{Fig. 2}       
\end{figure}

We have performed a set of single- and multi-polynomial fits with
$s_0\in [s_{min},s_{max}]$, where $s_{max}=3.0574 \mathrm{GeV}^2$ is
the maximum $s$ of the new spectral function combination. Fig.~\ref{Fig. 2}
(left panel) shows the $s_{min}$ dependence for the simplest fit, using
$w_0(x)=1$. We see that $\alpha_s(m_\tau )$ drifts at low $s_{min}$, where
the large-$s$ DV parametrization (\ref{DVs}) no longer describes the data
well for all $s>s_{min}$, but reaches a plateau (in the region of the yellow band)
as $s_{min}$ is increased, indicating that above such $s_{min}$ the form
(\ref{DVs}) is compatible with the underlying data. The value of
$\alpha_s(m_\tau )$ remains stable at larger $s_{min}$, with an error
that grows as the fit runs out of experimental data.

The right panel of Fig.~\ref{Fig. 2} shows results for $\alpha_s(m_\tau)$
and the $D=6$ condensate, $c_6$, obtained from one- or two-weight fits
using the weights listed in column 1. The values of both $\alpha_s(m_\tau )$
and $c_6$ are seen to be consistent across fits. Other fits were carried
out, always showing consistent results for $\alpha_s(m_\tau )$, the
condensates, and the DV parameters $\a,\b,\g$ and $\d$ \cite{Boito:2020xli}.
Combining this information, we find, in FOPT,
\beq
\lbl{alphas}
\alpha_s(m_\tau )=0.3077\pm 0.0065_{stat}\pm 0.0038_{pert.th.+DVs}
\Leftrightarrow \alpha_s(M_Z)=0.1171\pm 0.0010\  (\overline{MS}_{n_{f}=5})\ .
\eeq
where the first error is statistical and the second reflects perturbative
and DV uncertainties \cite{Boito:2020xli}. The result is in excellent
agreement with the PDG world average $\a_s(M_Z)=0.1179\pm 0.0009$
\cite{Workman:2022ynf}.

\section{The criticism of Pich and Rodriguez Sanchez}
\label{criticism}

In a 25th-anniversary special issue of the JHEP journal, the authors of
Ref.~\cite{Pich:2022tca} took the opportunity to express what they consider
to be the drawbacks of the approach  presented here, while arguing for the alleged
virtues of their own truncated OPE (tOPE) approach. Since the limited space
available here precludes refuting and/or responding in full to all the claims made there, we restrict ourselves to a couple of important clarifying remarks.\footnote{We will respond in full detail to all claims in a future regular
publication.}

\subsection{Comparing theory with experiment:   don't ignore correlations!}
\label{correlations}

The approach followed in Ref.~\cite{Pich:2022tca} fits $\alpha_s$ and
lower-$D$ OPE condensates using sets of higher-degree, at least doubly pinched,
polynomial FESRs, neglecting DVs, and, more importantly, neglecting higher-$D$
OPE condensate contributions in principle present for the weights
included in the analysis (cfr. our first message). The latter neglect is
often argued to be justified by the expectation that OPE contributions will
behave as $\sim C_D \left(\L_{QCD}^{2}/m_\t^{2}\right)^D$, where $\L_{QCD}$
is a typical QCD scale, of the order a few hundred MeV, and $C_D\sim 1$. This
argument, however, tacitly treats the OPE as if it were convergent. In fact,
the expansion is at best asymptotic, and the $C_D$ are expected to eventually
grow close to factorially with increasing $D$. Still, even if not justified, it might be that the tOPE assumption happens to work at the $\tau$ scale.

Let us look at some results purporting to support this suggestion, obtained
from analyses of $V+A$ $\tau$ spectral data in Ref.~\cite{Pich:2022tca}.
Table 2 of that reference quotes the CIPT results
$\alpha_s(m_\tau )=0.314^{+0.013}_{-0.009}$ and
$\alpha_s(m_\tau )=0.348^{+0.014}_{-0.012}$, obtained from the
$w(x)=1-2 x + x^2$ and $w(x)=1-x^6$ FESRs, in both cases setting
\emph{all} non-perturbative (NP) OPE and DV contributions to zero, even though
the former receives OPE contributions up to $D=6$ and the latter up to
$D=14$. These results (together with other entries in the table) are
characterized by the authors as showing ``amazing stability''. One should,
however, bear in mind that these values are extracted from the same
$\tau$ data and are \emph{very} strongly correlated. Declaring two results
$A$ and $B$ for the same observable to be compatible based on the overlap
of the two errors, $\delta A$ and $\delta B$, is extremely misleading
when the results are strongly correlated. A proper compatibility assessment
requires determining whether the difference $\Delta_{AB}=A-B$ is compatible
with zero within error, taking all correlations into account. When $A$ and
$B$ are essentially 100\% correlated, the error on $\Delta_{AB}$ is
$\delta_{AB}=|\delta A-\delta B|$ and, to claim compatibility, one should
find $\Delta_{AB}=0$ to within this correlated error $\d_{AB}$, not within
the sum $\delta A+\delta B$, which determines whether the two errors
overlap. The claimed compatibility of the two $\alpha_s$ values above
 is thus highly questionable (at best).\footnote{Assuming, e.g., a 90\% correlation, the above two values for $\a_s$ are 6 sigma apart.}

\begin{figure}[h]
\centering
\includegraphics[width=6cm,clip]{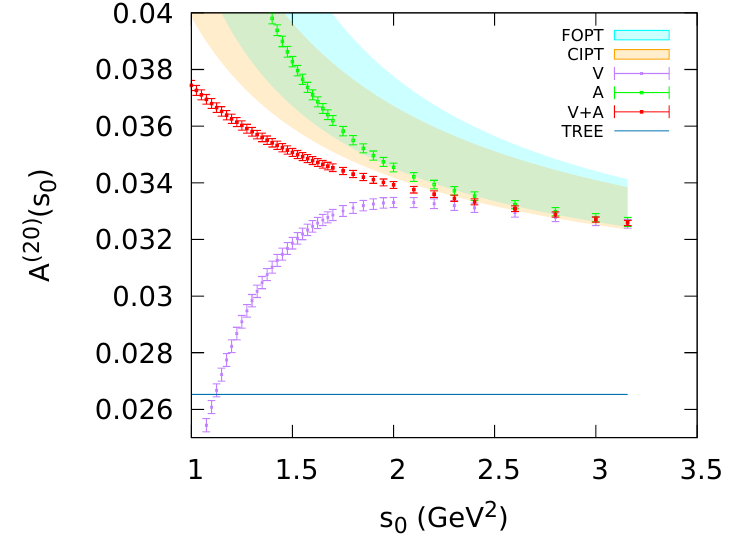}
\includegraphics[width=6cm,clip]{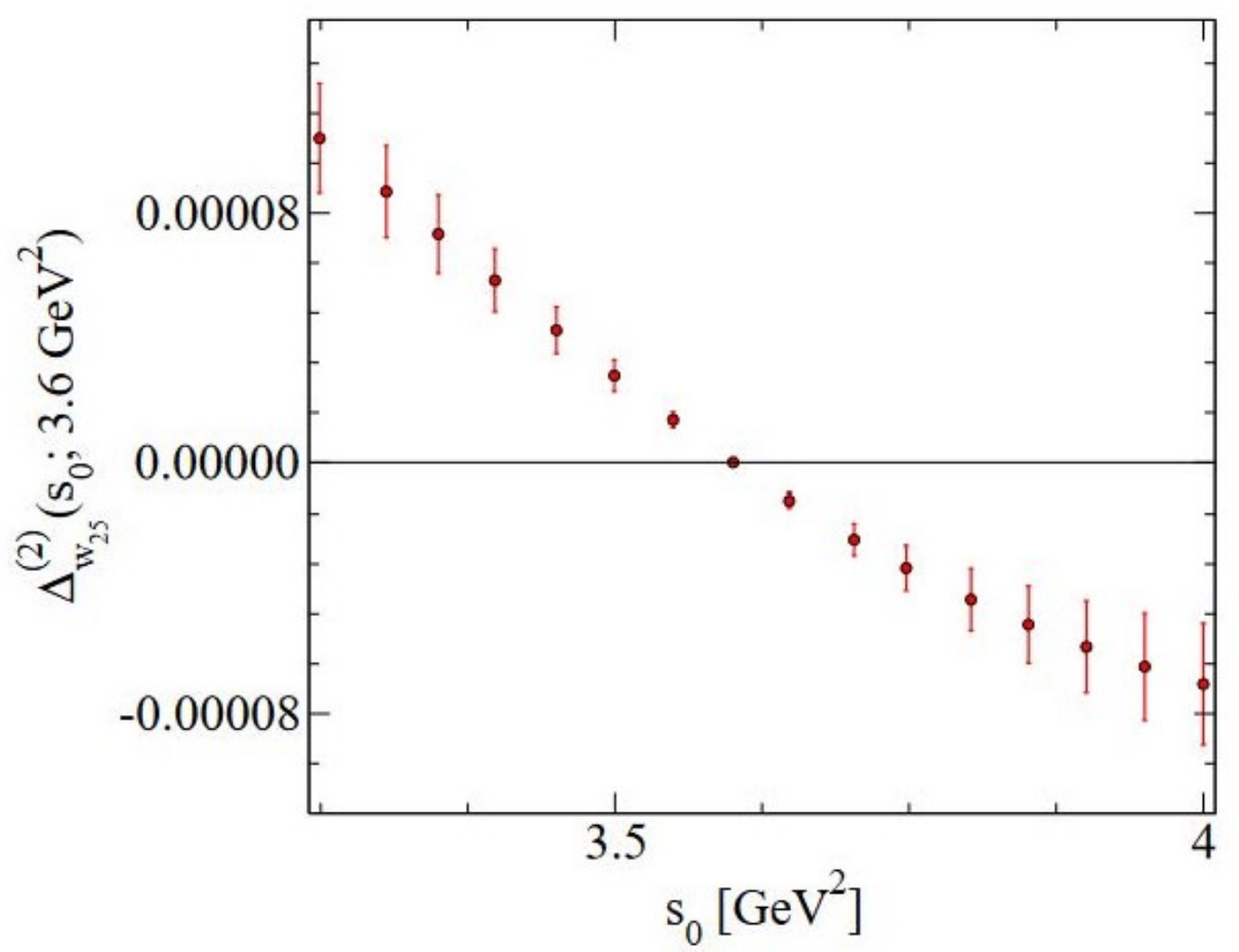}
\caption{Left panel: $s_0$ dependence of the LHS of the
$w^{(2,0)}(x)=(1-x)^2$ FESR for the $V$ (purple), $A$ (green)
and $\frac{1}{2}(V+A)$ (red) channels. The orange and
light-blue regions are the CIPT and FOPT perturbative predictions for
$\alpha_s(m_\tau )=0.329^{+0.020}_{-0.018}$. The blue horizontal line
at the bottom indicates the parton-model prediction. Taken from
Ref.~\cite{Pich:2022tca}. Right panel: Double difference
$\D^{(2)}_w(s_0,s_0^*)$ using $e^+e^-$ data (see text) .}
\label{Fig. 3}       
\end{figure}

Similar comments pertain to conclusions drawn from some of the
figures in Ref.~\cite{Pich:2022tca}, e.g., the left panel of
Fig.~\ref{Fig. 3}, which shows the $s_0$-dependence of the LHSs of the
$w^{(2,0)}(x)=(1-x)^2$ FESRs, Eq.~(\ref{FESR}), for the $V$ (purple),
$A$(green) and $\frac{1}{2}(V+A)$ (red) channels, together with
predictions obtained including perturbative CIPT and FOPT contributions
only, with fitted $\alpha_s$ as input, once more setting all NP
contributions to zero. The CIPT and FOPT results are shown, respectively,
by the orange and light-blue bands. From these curves the authors conclude
that this OPE truncation is confirmed by the data, since ``above $2.2\, \mathrm{GeV}^2$ all experimental
curves remain within the $1\sigma$ perturbative bands''. It is, however,
clear that the ``theory error'', represented by the orange and light-blue
bands, results largely from the errors in the data also
being shown. The declared agreement between theory and experimental
integrals involves a dangerous double counting of errors which
hides the impact of the very strong correlations between theory results
at different $s_0$, experimental results at different $s_0$, as well
as between the theory curves and underlying experimental data. As in our
previous example, the difference, in this case $theory - data$, employing
an error assessment which takes into account all of these correlations, would
provide a more reliable means of assessing the supposed compatibility.
Regrettably, such checks  are not shown in Ref.~\cite{Pich:2022tca}.

The authors of Ref.~\cite{Pich:2022tca} have also argued that the existence of acceptable
fits for 4 OPE parameters ($\alpha_s$ and either the $D=4,6$ and $8$
or $D=6,8$ and $10$ condensates) using 5 independent $s_0=m_\tau^2$
FESRs with at-least-doubly-pinched weights of degree up to 7 (and hence,
in principle, OPE contributions up to $D=16$) establishes the validity of
the neglect of in-principle-present $D=10, 12, 14$ and $16$ or $D=12, 14$
and $16$ contributions. They have, moreover, argued against using the $s_0$
dependence of the resulting fits at lower $s_0$ to test this assumption, on
the grounds that considering such lower-$s_0$ input enhances neglected DV
contributions. While kinematic restrictions preclude carrying out $\tau$-based
analyses at higher $s_0$, where this rational for avoiding $s_0$-dependence
tests would not apply, one can perform such higher-$s_0$, tOPE assumption
tests by shifting to analyses of $e^+ e^-$ data, where $s_0$ is kinematically
unrestricted.
According to the arguments of
Ref.~\cite{Pich:2022tca}, if a 5-weight, 4-OPE-parameter tOPE fit to
$e^+ e^-$ hadroproduction data is successful at some sufficiently large
$s_0$, $s_0^*$, then, since NP OPE condensate and DV
contributions are supposedly negligible at $s_0=s_0^*$, they will be
even more negligible at even higher $s_0$ and the tOPE fit results will
continue to provide a good representation of the experimental spectral
integrals of the analysis at $s_0>s_0^*$. As noted above, to draw
statistically meaningful conclusions, it is necessary to take all
correlations into account, including those between data and the
fitted OPE parameters. In Ref.~\cite{Boito:2016oam} we carried out
this test, first establishing that an acceptable 5-weight, 4-OPE-parameter
tOPE fit indeed existed, for $s_0^*=3.6\ {\rm GeV}^2$, and then evaluating
the double differences between the LHSs and RHSs of Eq.~(\ref{FESR}),
$\D^{(2)}_w(s_0,s_0^*)=\left[I_w^{th}-I_w^{exp}\right](s_0)-
\left[I_w^{th}-I_w^{exp}\right](s_0^*)$, produced by this fit
at $s_0>s_0^*$. The right panel in Fig.~\ref{Fig. 3} shows an example
of the results obtained, in this case for one of the weights,
$w_{25}(x)=1-7 x^6+6 x^7$, which forms part of the ``optimal weight"
set of Ref.~\cite{Pich:2016bdg}. It is clear that, away from $s_0 = s_0^*$, $\D^{(2)}_w(s_0,s_0^*)$
is nowhere near zero within errors. See  Ref.~\cite{Boito:2016oam} for more
details and further examples. It is clear that the tOPE assumption fails
badly, and this leads us to our third message:

\vspace{.1cm}
\emph{To establish agreement between two strongly correlated observables
$A$ and $B$, the result for $A-B$ should be shown to be 0 within errors
obtained taking  all correlations into account.}

\subsection{Duality Violations}
\label{duality}

Though the DV parametrization (\ref{DVs}) is supposed to be valid up to
corrections falling off at large $s$, the authors of Ref.~\cite{Pich:2016bdg}
have performed analyses with this expression multiplied by a factor $s^k$,
with $k$ a \emph{positive} integer as large as $8$, and concluded that the
result for $\alpha_s(m_\tau )$ is very sensitive to $k$ and hence to the
form chosen for the DV contribution to the spectral function.\footnote{Even though a similar spread in $\a_s$ values is found in tests of the tOPE strategy.} Interestingly,
the same authors made no attempt to study the effect of adding such an $s^k$
factor in a whole series of papers analyzing the $V-A$ correlator using
precisely the DV parametrization shown in Eq.~(\ref{DVs}) (see Ref.~\cite{Gonzalez-Alonso:2016ndl} and references therein). Such a factor, in any case, makes little sense
given the functional dependence of the corrections to the large-$s$
limit indicated in Eq.~(\ref{DVs}).

\begin{figure}[h]
\centering
\includegraphics[width=5cm,clip]{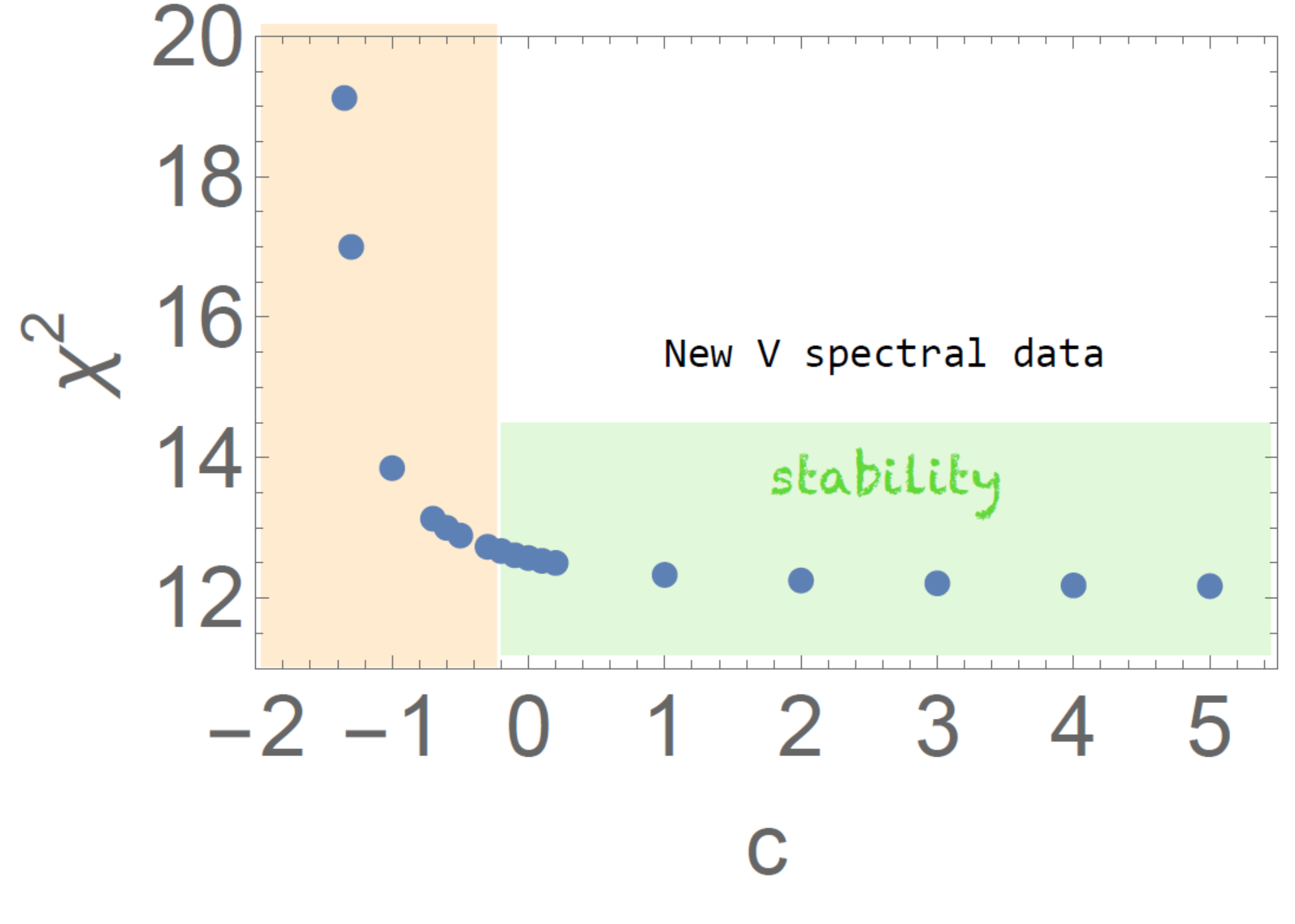}\hspace{.2cm}
\includegraphics[width=5cm,clip]{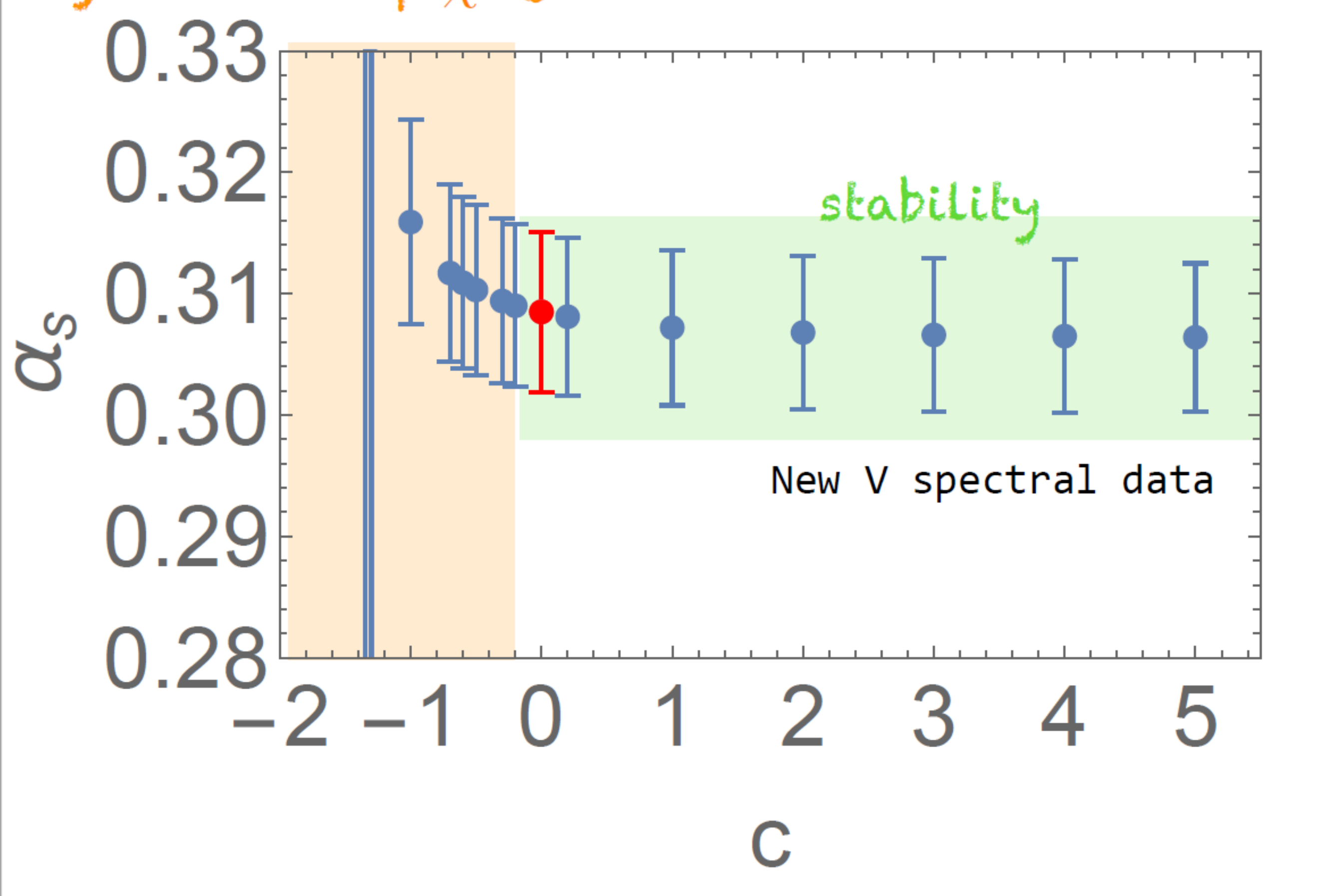}
\caption{$\chi^2$ (left panel) and  $\alpha_s(m_\tau )$ (right panel) as functions of $c$ (see text) from a fit to the new $V$-channel tau data. The red point is our result for $w_0=1$ in the RHS panel of
Fig.~\ref{Fig. 2}.}
\label{Fig. 4}       
\end{figure}

\vspace{-.2cm}

A variation which does make sense is obtained by multiplying the expression
(\ref{DVs}) by a correction $(1+c/s)$, with $c$ a constant \cite{Boito:2017cnp}, which the authors
of Ref.~\cite{Pich:2022tca} also explored, obtaining the results
$\alpha_s(m_\tau )=0.319$ and $\alpha_s(m_\tau )=0.260$ for the choices
$c=-1.35$ and $c=-2$ GeV$^2$, in fits with $s_{min}=1.55\, \mathrm{GeV}^2$ to the
$w_0=1$, ALEPH $V$-spectral-function-based \cite{Davier:2013sfa} FESR.
Looking at the difference between the two central values, the authors of
this reference concluded that the DV parametrization (\ref{DVs}) induces
a large systematic uncertainty in the determination of $\alpha_s$.
Although not quoted in Ref.~\cite{Pich:2022tca}, the errors on these
results turn out to be $\pm 0.016$ and $\pm 0.089$, respectively.
The ALEPH data employed is thus, in fact, insufficiently precise
to allow such a conclusion to be drawn. Here, the new improved V
channel data can help \cite{Boito:2020xli}. The left panel of
Fig. \ref{Fig. 4} shows the $\chi^2$ results for the same fit, now using the improved V data, as a function of the correction parameter $c$. One sees a very steep increase in $\chi^2$
for negative $c$, in particular for the values $c=-1.35$ and $c=-2$ GeV$^2$
chosen in Ref.~\cite{Pich:2022tca}. These two values, therefore, should
be considered unphysical. In addition, even with the improved data, the
flatness of the $\chi^2$ at $c>0$ means the fit is unable to determine
a preferred value for $c$. The right panel of Fig. \ref{Fig. 4} shows
the $c$-dependence of the $\alpha_s(m_\t)$ obtained from this fit.
Although one expects a small value of $c$ to be most reasonable, even
for large $c>0$, where the term  $(1+c/s)$ could no longer be considered
a correction, e.g., in the region $s\sim 2\, \mathrm{GeV}^2$, the result for
$\a_s(m_\t)$ is very insensitive to this term because its effect gets
absorbed in the DV parameters. The conclusion is that the error quoted for our
determination of $\alpha_s(m_\tau )$ in Eq. (\ref{alphas}) is reliable,
and already includes a sensible estimate for the systematic error
coming from the DV parametrization, contrary to the claim made in
Ref.~\cite{Pich:2022tca}. We finally conclude with our last message:

\vspace{0.2cm}
\emph{Don't ignore Duality Violations
if you want to learn about their impact on $\alpha_s$ from FESRs.}

\vspace{0.5cm}

\noindent
{\bf Acknowledgements}: DB is supported by the S\~{a}o Paulo Research Foundation (FAPESP) No.
2021/06756-6, by CNPq Grant No. 308979/2021-4. MG is and WS was supported
by the U.S. Department of Energy, Office of Science, Office of High Energy
Physics, under Award No. DE-SC0013682. MVR was supported by FAPESP grant
No. 2019/16957-9.  KM is supported by a grant from the Natural Sciences
and Engineering Research  Council of Canada. SP is supported by the
Spanish Ministry of Science, Innovation and Universities (project
PID2020-112965GB-I00/AEI/10.13039/501100011033)  and by Grant 2017 SGR 1069.
IFAE is partially funded by the CERCA program of the Generalitat de Catalunya.
\bibliography{references}

 \end{document}

%% file: defs.tex
\let\ifcomments\iftrue
\def\commentsoff{\global\let\ifcomments\iffalse}

\let\commentsize\small
\def\tinycomments{\global\let\commentsize\footnotesize}

\def\nolabel{???}
\def\printmylabel{\nolabel}
\let\ifshowlabels\iffalse

\catcode`\@=11
\def\showlabels{
  \global
  \def\@eqnnum{{\normalfont \normalcolor (%
    \theequation%
    \hspace{+3ex}%
    \framebox[10ex]{{\scriptsize \printmylabel}}
    \hspace{-13ex}
  )}}
  \global\let\ifshowlabels\iftrue
}

\catcode`\@=12

\def\lbl#1{
  \label{#1}
  \global\def\printmylabel{#1}
}

\def\beq{
  \def\printmylabel{\nolabel}         
  \begin{equation}
}

\def\bqry{
  \def\printmylabel{\nolabel}
  \begin{eqnarray}
}

\def\eeq{\end{equation}}

\def\eqry{\end{eqnarray}}

\def\eqsbysecs{}
\newif\ifeqsbysecs

\newif\ifEqsBySubsecsCurrent

\newif\ifEqsBySubsecsActive
\def\setEqsBySubsecsActive{
  \ifEqsBySubsecsCurrent
    \global\EqsBySubsecsActivetrue
  \else
    \global\EqsBySubsecsActivefalse
  \fi}

\newcounter{lemma}

\newcounter{secname}
\newcounter{subsecname}
\newcounter{subsubsecname}

\newenvironment{secname}{%
  \refstepcounter{secname}%
  \setcounter{subsecname}{0}%
  \ifeqsbysecs%
    \setcounter{equation}{0}%
    \setcounter{lemma}{0}%
    \global\def\eqsbysecs{\arabic{secname}.}%
    \setEqsBySubsecsActive%
    \ifEqsBySubsecsActive%
      \global\def\eqsbysecs{\arabic{secname}.\arabic{subsecname}.}%
    \fi%
  \fi%
}{}

\newenvironment{subsecname}{%
  \refstepcounter{subsecname}%
  \setcounter{subsubsecname}{0}%
  \ifEqsBySubsecsActive%
    \setcounter{equation}{0}%
  \fi%
}{}

\newenvironment{subsubsecname}{%
  \refstepcounter{subsubsecname}%
}{}

\newcounter{appname}
\newcounter{subappname}
\newcounter{subsubappname}

\newenvironment{appname}{%
  \refstepcounter{appname}%
  \setcounter{subappname}{0}%
  \ifeqsbysecs
    \setcounter{equation}{0}
    \setcounter{lemma}{0}
    \global\def\eqsbysecs{\Alph{appname}.}
  \fi
}{}

\newenvironment{subappname}{%
  \refstepcounter{subappname}%
  \setcounter{subsubappname}{0}%
}{}

\newenvironment{subsubappname}{%
  \refstepcounter{subsubappname}%
}{}

\newcounter{subeq}

\newenvironment{lemma}{%
  \refstepcounter{lemma}%
  }{}

\def\a{\alpha}
\def\b{\beta}

\def\d{\delta}
\def\g{\gamma}

\def\i{\iota}

\def\m{\mu}
\def\n{\nu}

\def\p{\pi}                     
\def\s{\sigma}                  
\def\t{\tau}

\def\D{\Delta}

\def\L{\Lambda}

\def\P{\Pi}

\def\cbo{{\,\raise-.15ex\Sc [\,}}                       

\def\ddt#1{{\buildrel {\hbox{\LARGE .\kern-2pt.}} \over {#1}}}

